\documentclass[amsmath,amssymb,superscriptaddress,nobalancelastpage,aps,prl,twocolumn]{revtex4-1}

\usepackage[utf8]{inputenc}
\usepackage{graphicx}
\usepackage{bm}
\usepackage{amsmath}
\usepackage{amssymb}
\usepackage{hyperref}
\usepackage{subfigure}
\usepackage{float}
\usepackage{multirow}
\usepackage{booktabs}
\usepackage{varioref}
\usepackage{xr-hyper}
\usepackage[dvipsnames]{xcolor}
\usepackage{nicefrac}
\usepackage{xfrac}
\usepackage{hyperref}
\hypersetup{colorlinks,linkcolor=blue,urlcolor=blue,citecolor=blue}
\usepackage[normalem]{ulem}
\usepackage{siunitx}
\usepackage{graphicx}
\usepackage{dcolumn}
\usepackage{braket}
\usepackage{wasysym}
\usepackage{textcomp}
\usepackage{upgreek}
\usepackage{setspace}
\usepackage{lineno}
\usepackage{autobreak}
\usepackage{lipsum}

\begin{document}
\setstretch{1.1}

\title{$g$-wave altermagnetic order parameter in hematite}

\author{Tianren~Wang}
\affiliation{Department of Physics, The Chinese University of Hong Kong, Shatin, Hong Kong, China}

\author{Yuehong~Li}
\affiliation{Department of Physics, The Chinese University of Hong Kong, Shatin, Hong Kong, China}

\author{Yu~Feng}
\email{fengyu@ihep.ac.cn}
\affiliation{Institute of High Energy Physics, Chinese Academy of Sciences, Beijing 100049, China}
\affiliation{Spallation Neutron Source Science Center, Dongguan 523803, China}

\author{Andong~Liu}
\affiliation{Department of Physics, The Chinese University of Hong Kong, Shatin, Hong Kong, China}

\author{Yuetong~Wu}
\affiliation{Department of Physics, The Chinese University of Hong Kong, Shatin, Hong Kong, China}

\author{Qian~Zhao}
\affiliation{Institute of High Energy Physics, Chinese Academy of Sciences, Beijing 100049, China}
\affiliation{Spallation Neutron Source Science Center, Dongguan 523803, China}

\author{Yujie~Yan}
\affiliation{Department of Physics, The Chinese University of Hong Kong, Shatin, Hong Kong, China}

\author{Wei~Luo}
\affiliation{Institute of High Energy Physics, Chinese Academy of Sciences, Beijing 100049, China}
\affiliation{Spallation Neutron Source Science Center, Dongguan 523803, China}

\author{Xin~Tong}
\affiliation{Institute of High Energy Physics, Chinese Academy of Sciences, Beijing 100049, China}
\affiliation{Spallation Neutron Source Science Center, Dongguan 523803, China}
\affiliation{Guangdong Provincial Key Laboratory of Extreme Conditions, Dongguan, 523803, China}

\author{Yi~Lu}
\affiliation{National Laboratory of Solid State Microstructures and
Department of Physics, Nanjing University, Nanjing, China}
\affiliation{Collaborative Innovation Center of Advanced Microstructures, Nanjing, China}

\author{Yao~Shen}
\affiliation{Beijing National Laboratory for Condensed Matter Physics, Institute of Physics, Chinese Academy of Sciences, Beijing 100190, China}
\affiliation{School of Physical Sciences, University of Chinese Academy of Sciences, Beijing 100049, China}

\author{Stefano~Agrestini}
\affiliation{Diamond Light Source, Harwell Campus, Didcot, Oxfordshire OX11 0DE, United Kingdom}

\author{Jaewon~Choi}
\email{jaewon.choi@snu.ac.kr}
\affiliation{Department of Physics and Astronomy, Seoul National University, Seoul 08826, Republic of Korea}

\author{Qisi~Wang}
\email{qwang@cuhk.edu.hk}
\affiliation{Department of Physics, The Chinese University of Hong Kong, Shatin, Hong Kong, China}
\affiliation{State Key Laboratory of Quantum Information Technologies and Materials, The Chinese University of Hong Kong, Shatin, Hong Kong, China}


\begin{abstract}
Altermagnets combine the vanishing net magnetization of antiferromagnets with momentum-dependent spin splitting. Magnon band splitting provides a direct probe of altermagnetic order and may enable chirality-selective magnon transport, yet the momentum-space symmetry of this splitting has not been determined quantitatively.
Here we use inelastic neutron scattering to map the momentum dependence of altermagnetic magnon splitting in hematite ($\alpha$-Fe$_2$O$_3$).
The splitting vanishes along nodal directions and reaches maxima off the nodes, revealing the $g$-wave symmetry of the altermagnetic order parameter.
These results agree with linear spin-wave theory calculations based on the altermagnetic model, which further identify the nondegenerate branches as magnons of opposite chirality and trace the splitting to symmetry-inequivalent long-range exchange interactions. Our results provide the first quantitative determination of the momentum-space symmetry of altermagnetic chiral magnons. These findings, together with hematite's high magnetic ordering temperature and low magnon damping, establish it as a promising platform for low-dissipation, symmetry-selective magnonic applications.
\end{abstract}

\maketitle

Altermagnetism has recently emerged as a distinct class of collinear compensated magnetic states, complementary to conventional ferromagnetism and antiferromagnetism~\cite{smejkal_2022_conventional,smejkal_2022_emerging,jungwirth_2026_symmetry,cheong_2025_altermagnetism}. In altermagnets, the opposite-spin sublattices are related by crystal rotations or mirror operations, in contrast to conventional antiferromagnets, in which the opposite-spin sublattices are connected by inversion or translation symmetry~\cite{smejkal_2022_conventional,smejkal_2022_emerging,liu_2022_spin}. This symmetry condition allows a vanishing net magnetization to coexist with exchange-scale spin splitting in momentum space~\cite{smejkal_2022_conventional, smejkal_2022_emerging}. Experimentally, altermagnetic (AM) spin splitting has been reported in several materials, ranging from semiconducting MnTe~\cite{krempasky_2024_altermagnetic, lee_2024_mnte} and MnTe$_2$~\cite{zhu_observation_2024}, to metallic CrSb~\cite{reimers_2024_direct,ding_2024_large,yang_2025_crsb} and KV$_2$Se$_2$O~\cite{jiang_2025_metallic}. A central challenge in this field is to experimentally determine not only whether a material is altermagnetic but also the symmetry of its altermagnetic order parameter. As in unconventional superconductors, where the momentum-dependent gap function defines the symmetry of the superconducting state, the altermagnetic order parameter can be expressed through the momentum-dependent band splitting. Its nodal structure and sign reversals distinguish $d$-, $g$-, and higher-order altermagnetic phases~\cite{long_2026_crsb}. 
In magnetic insulators, electronic techniques such as angle-resolved photoemission spectroscopy (ARPES) cannot readily access low-energy quasiparticles, whereas magnon spectroscopy probes the relevant spin excitations directly~\cite{Maier_2023_Weakcoupling,Liu2024Chirala,sun_observation_2025}.

\begin{figure*}[t]
    \centering
    \includegraphics[width=0.95\linewidth]{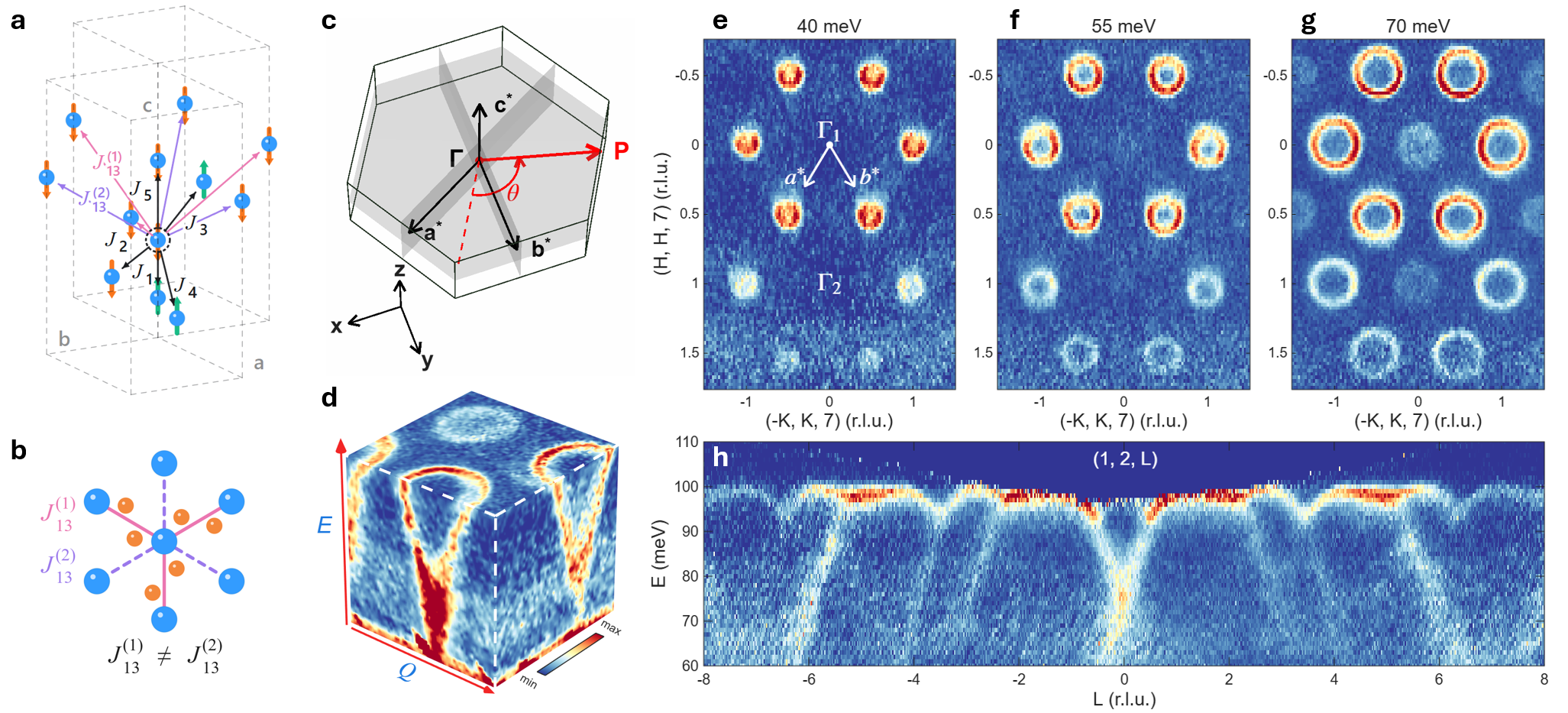}
    \caption{\textbf{Magnetic interactions and excitations in $\alpha$-Fe$_2$O$_3$.} (a) Exchange paths around a representative Fe site. The dominant $J_1$--$J_5$ interactions are shown in black, and the inequivalent long-range $J_{13}^{(1)}$ and $J_{13}^{(2)}$ paths in pink and purple. Orange and green arrows mark the oppositely oriented magnetic sublattices.
    (b) Top-view schematic of the microscopic origin of the altermagnetic magnon splitting. Distinct oxygen-mediated environments make $J_{13}^{(1)}$ stronger than $J_{13}^{(2)}$, lifting the degeneracy between the $S^{+-}$ and $S^{-+}$ magnon branches.
    (c) Brillouin-zone geometry, with $\mathbf{P}$ denoting the in-plane propagation direction and $\theta$ its azimuth relative to the dashed off-nodal direction.
    (d) Quasi-three-dimensional INS intensity in momentum--energy space along $(H,0,0)$ and $(-0.5K,K,0)$.
    (e-g) Constant-energy slices in the $(H,H,0)$ and $(-K,K,0)$ planes, showing expanding ring-like contours around zone centres $\Gamma_1$ and $\Gamma_2$.
    (h) INS spectrum along $L$ through $\mathbf{Q}=(1,2,L)$, showing dispersing magnon branches.}
    \label{fig:overall}
\end{figure*}

Among candidate altermagnets, hematite, $\alpha$-Fe$_2$O$_3$, combines a high N\'{e}el temperature ($T_N=950$~K), insulating character, ultralow magnetic damping, and exceptional chemical stability~\cite{dannegger_2023_fe2o3,hamdi_2023_fe2o3,lebrun_2020_fe2o3}. 
It also undergoes the Morin transition near $T_M \approx 260$~K, a spin-reorientation transition that changes the N\'{e}el-vector orientation and magnetic anisotropy, thereby providing a means to control magnon polarization and transport and making hematite particularly attractive for antiferromagnetic spintronics~\cite{lebrun_2020_fe2o3,morin_1950_fe2o3,dannegger_2023_fe2o3}.
Recent theoretical work has shown that the crystal and magnetic symmetries of hematite support $g$-wave altermagnetism associated with high-rank magnetic multipolar order~\cite{verbeek_2024_fe2o3}. First-principles calculations further predict that this order produces a strongly direction-dependent splitting of oppositely polarized magnon branches, while relativistic effects, including Dzyaloshinskii--Moriya interactions and magnetic anisotropy, do not mask the dominant nonrelativistic altermagnetic splitting away from the Brillouin-zone centre~\cite{hoyer_2025_fe2o3}.

\begin{figure*}[t]
    \centering
    \includegraphics[width=0.98\linewidth]{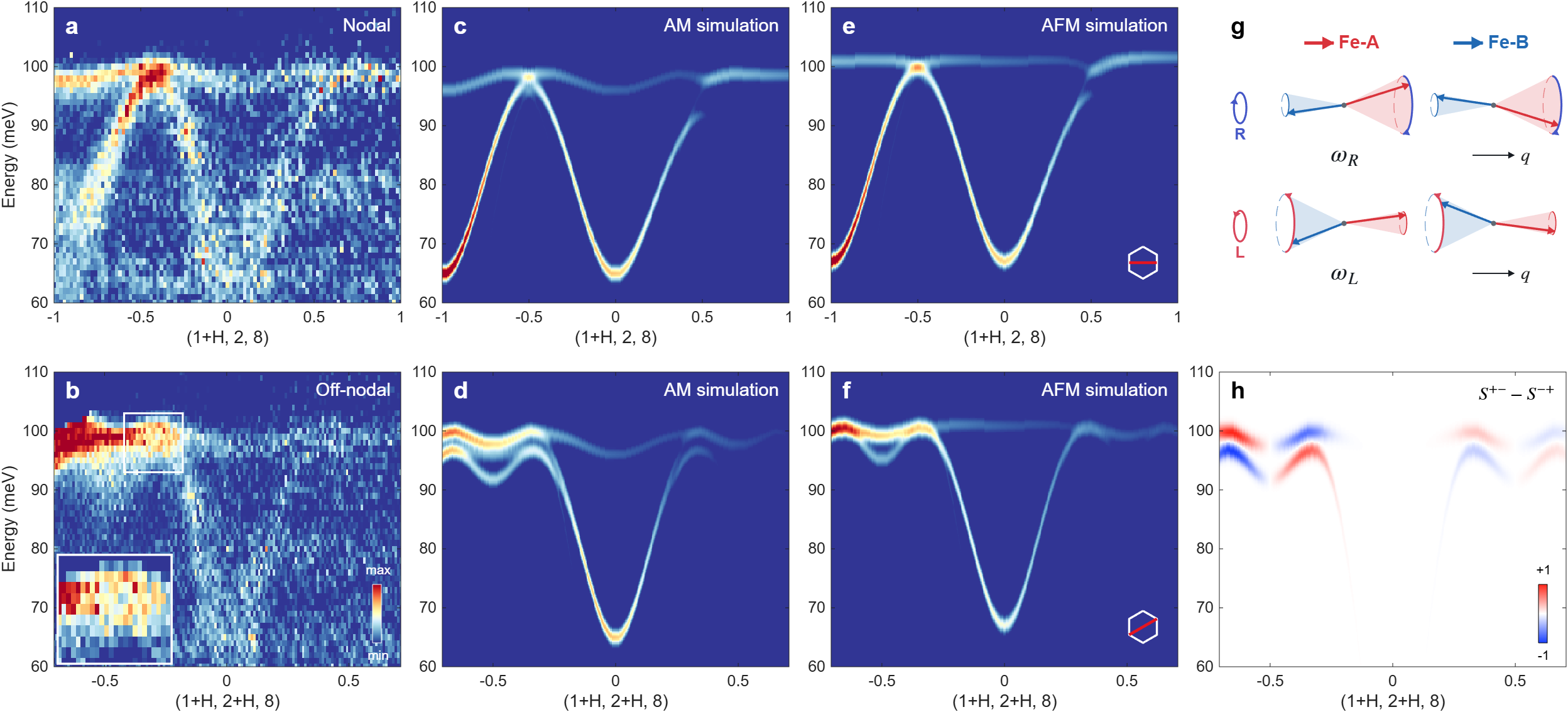}
    \caption{\textbf{Magnon dispersion and its chiral splitting.} (a,b) Experimental energy-momentum spectra along the nodal direction, $\mathbf{Q}=(1+H,2,8)$, and the off-nodal direction, $\mathbf{Q}=(1+H,2+H,8)$, respectively. The inset in (b) highlights the high-energy branch splitting. 
    (c-f) LSWT calculations for the AM model (c,d) and the corresponding conventional AFM model (e,f). (g) Semiclassical schematic of right- and left-handed magnon eigenmodes propagating along $\mathbf{q}$. Each row depicts a single mode involving both Fe sublattices; red and blue arrows denote the Fe$_\mathrm{A}$ and Fe$_\mathrm{B}$ moments, respectively, and the cones indicate their precession, with handedness defined relative to $+\mathbf{q}$. Altermagnetic exchange lifts the degeneracy, giving $\omega_R(\mathbf{q})\neq\omega_L(\mathbf{q})$. Panel (h) displays the calculated chiral response, $S^{+-}(\mathbf{Q},\omega)-S^{-+}(\mathbf{Q},\omega)$, where red and blue denote excitations with opposite magnon chiralities. Intensities are given in arbitrary units.}
    \label{fig:dispersion}
\end{figure*}

Experimental evidence for altermagnetic magnon splitting in hematite has recently been reported by inelastic neutron scattering (INS)~\cite{sun_observation_2025}. A magnon band splitting of approximately 3~meV was observed between the optical branches near 100~meV, attributed to alternating exchange interactions between symmetry-inequivalent thirteenth-neighbour Fe pairs as predicted by theoretical proposals of altermagnetic phase in this compound~\cite{hoyer_2025_fe2o3}. 
However, the splitting was resolved only along selected momentum trajectories, leaving its in-plane angular dependence undetermined. Such momentum-space information is essential because the in-plane angular variation of the chiral magnon splitting, \mbox{$|\Delta(\mathbf{Q})|=|E_{\circlearrowleft}(\mathbf{Q})-E_{\circlearrowright}(\mathbf{Q})|$}, directly encodes the altermagnetic order parameter. In particular, the positions of symmetry-protected nodes and antinodes, together with the characteristic angular structure of the splitting magnitude, distinguish $g$-wave altermagnetism from other symmetries and conventional magnon splitting arising from relativistic or dipolar interactions~\cite{verbeek_2024_fe2o3,hoyer_2025_fe2o3, Faure_2025_mnf2}.

Here we report an inelastic neutron scattering study of single-crystalline $\alpha$-Fe$_2$O$_3$, focusing on the in-plane momentum dependence of the optical magnon branches. By mapping the angular modulation of the magnon splitting, we identify the nodal and off-nodal directions and resolve its $g$-wave pattern. Combined with linear spin-wave theory (LSWT) calculations which determine the alternating chiral character of the split branches, these results unambiguously establish the $g$-wave altermagnetic magnon in hematite. We further show that the symmetry-inequivalent long-range exchange interactions responsible for the splitting also modify the magnon eigenvectors, redistributing spectral weight between the magnon branches.
The comprehensive analysis provides stringent evidence for anisotropic exchange in the altermagnetic phase and distinguishes its contribution from relativistic anisotropy effects. 
By resolving the momentum dependence of the chiral magnons splitting, we identify key propagation directions for future magnon-transport experiments and symmetry-selective magnonic applications~\cite{lebrun_2018_fe2o3,lebrun_2020_fe2o3}.\\

\noindent\textbf{Results}\\
Figure~\ref{fig:overall}(a) illustrates the exchange network around a representative Fe site in $\alpha$-Fe$_2$O$_3$. Of particular importance are the two symmetry-inequivalent thirteenth-neighbour exchange pathways, $J_{13}^{(1)}$ and $J_{13}^{(2)}$ (Figure~\ref{fig:overall}(b)). Theoretically, this exchange asymmetry generates a momentum-dependent exchange field that lifts the degeneracy of oppositely polarized magnon branches, providing the microscopic basis for the altermagnetic splitting. Its in-plane angular dependence (Fig.~\ref{fig:overall}(c)), reflects the order parameter symmetry~\cite{sun_observation_2025,hoyer_2025_fe2o3}.

An overview of the measured scattering intensity is displayed in Fig.~\ref{fig:overall}(d) as a quasi-three-dimensional rendering in $I(E,\mathbf{Q})$ space, revealing well-defined dispersive magnon modes consistent with previous reports~\cite{sun_observation_2025,samuelsen_1970_fe2o3}. Representative constant-energy slices in the $(H,K,7)$ plane and the dispersion along the out-of-plane $L$ direction through $\mathbf{Q}=(1,2,L)$ are shown in Fig.~\ref{fig:overall}(e-g) and Fig.~\ref{fig:overall}(h), respectively. Together, these data capture the three-dimensional spin-wave excitations, comprising steep branches emerging from successive zone centres and weakly dispersive optical modes near 100~meV.

\begin{figure*}[tbh]
    \centering
    \includegraphics[width=0.95\linewidth]{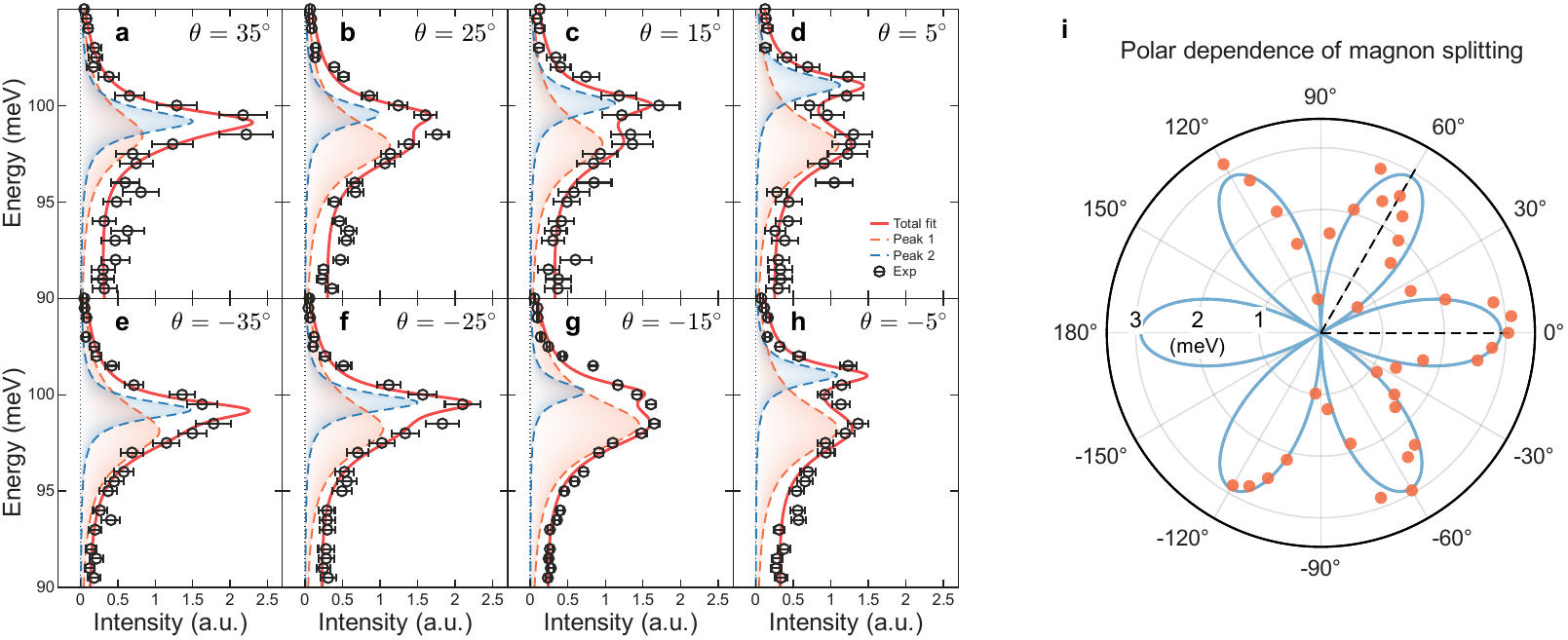}
    \caption{\textbf{Angular dependence of the magnon band splitting.} (a-h) Representative energy cuts at selected in-plane angles $\theta$ as indicated around $\mathbf{Q}=(1,2,8)$. Black circles show the experimental intensity. The red curves show the total fits, which consist of a linear background and two Lorentzian functions (blue and red areas). Error bars denote one standard deviation. a.u., arbitrary units. (i) Polar plot of the magnon splitting. Red dots represent the fitted splitting magnitudes from the energy cuts as in (a-h). The blue solid curve denotes calculated magnitude of the $g$-wave altermagnetic magnon splitting using LSWT.}
    \label{fig:polar}
\end{figure*}

To characterize the magnon dispersion, we trace the excitations along two high-symmetry momentum directions through the magnetic zone centre $\mathbf{Q}=(1,2,8)$ (Fig.~\ref{fig:dispersion}).
The large $L$ component ensures sufficient energy and momentum coverage to resolve the structure of any magnon splitting (see Fig.~\ref{fig:overall}(h)). Consistent observations are obtained in other Brillouin zones (Supplementary Information).
Fig.~\ref{fig:dispersion}(a) displays the dispersion along the $H$ direction, $\mathbf{Q}=(1+H,2,8)$, corresponding to a nodal direction of the predicted altermagnetic splitting. Along this trajectory, no resolvable branch separation is observed throughout the measured energy range. By contrast, a splitting emerges near 100~meV along the off-nodal $(H,H)$ direction, $\mathbf{Q}=(1+H,2+H,8)$ (Fig.~\ref{fig:dispersion}(b)).

To model this direction-dependent splitting, we performed LSWT calculations based on an effective Heisenberg Hamiltonian~\cite{sun_observation_2025}:
\begin{equation}
\mathcal{H}_{\text{AM}} = \mathcal{H}_\text{AFM}+\Delta \left(\sum_{\langle i,j \rangle_{13}^{(1)}}\mathbf{S}_i \cdot \mathbf{S}_j-\sum_{\langle i,j \rangle^{(2)}_{13}}\mathbf{S}_i \cdot \mathbf{S}_j \right)
\end{equation}

where

\begin{equation}
    \mathcal{H}_{\text{AFM}} =
    \sum_{\langle i,j\rangle_n} J_n \mathbf{S}_i \cdot \mathbf{S}_j
    + \sum_{\langle i,j\rangle_{13}} J' \mathbf{S}_i \cdot \mathbf{S}_j
    + \sum_i D(S_i^z)^2
\end{equation}

\begin{equation}
    \Delta=\frac{J_{13}^{(1)}-J_{13}^{(2)}}{2},  J'=\frac{J_{13}^{(1)}+J_{13}^{(2)}}{2}
\end{equation}
Here, $J_n$ denotes the $n$-th-neighbour exchange coupling ($n=1,2,3,4,5$), $\langle i,j\rangle$ labels the interacting spin pairs shown in Fig.~\ref{fig:overall}(a,b), and $\mathbf{S}_i$ is the spin operator at lattice site $i$. $D$ denotes the single-ion anisotropy, with $S_i^z$ the $z$ component of $\mathbf{S}_i$. 
Within this formulation, $\mathcal{H}_{\mathrm{AM}}$ describes the altermagnetic model with $J_{13}^{(1)} \neq J_{13}^{(2)}$, whereas $\mathcal{H}_{\mathrm{AFM}}$ describes the spin-degenerate AFM reference obtained by setting $\Delta=0$, for which $J_{13}^{(1)}=J_{13}^{(2)}=J'$.

Whereas the spin-degenerate AFM calculation fails to reproduce the observed off-nodal magnon splitting (Fig.~\ref{fig:dispersion}(f)), the AM calculation successfully captures this feature (Fig.~\ref{fig:dispersion}(d)).
The fitted exchange parameters using the AM model are $J_1=-1.51(4)$, $J_2=-0.45(3)$, $J_3=5.44(9)$, $J_4=4.05(4)$, $J_5=0.20(4)$, $J_{13}^{(1)}=0.321(3)$ and $J_{13}^{(2)}=0.097(3)$~meV. The single-ion anisotropy was fixed at $D=-0.02$~meV~\cite{sun_observation_2025}.
These parameters retain the previously reported exchange hierarchy, with the dominant $J_3$ and $J_4$ values close to the earlier estimates~\cite{sun_observation_2025}.

The chiral character of the calculated branches is illustrated in Fig.~\ref{fig:dispersion}(g,h). 
In a semiclassical picture, the split branches correspond to right- and left-handed magnon modes (Fig.~\ref{fig:dispersion}(g)). The altermagnetic order breaks time-reversal symmetry, and only time reversal combined with the crystal rotation relating the opposite-spin Fe sublattices connects the two modes at rotated momenta. Away from symmetry-protected nodal directions, their degeneracy at the same momentum is therefore not enforced, allowing $\omega_R(\mathbf{Q})\neq\omega_L(\mathbf{Q})$.
In $\alpha$-Fe$_2$O$_3$, this splitting arises from inequivalent $J_{13}^{(1)}$ and $J_{13}^{(2)}$ exchange paths with distinct oxygen-mediated environments, producing a $g$-wave modulation of the symmetric exchange field (Fig.~\ref{fig:overall}(b)). The calculated chiral response, $S^{+-}(\mathbf{Q},\omega)-S^{-+}(\mathbf{Q},\omega)$, shown in Fig.~\ref{fig:dispersion}(h), assigns opposite chiralities to the two off-nodal branches. Thus, although the present unpolarized INS measurement probes the sum of the two channels, the LSWT calculation identifies the split modes as magnons of opposite chirality.

Having established that the altermagnetic Hamiltonian captures the direction-dependent magnon spectra, we next extract the angular dependence of the magnon splitting that characterizes the altermagnetic order parameter.
Fig.~\ref{fig:polar}(a-h) show representative energy cuts at selected in-plane propagation angles $\theta$, with $\theta=0^\circ$ corresponding to one off-nodal direction. Each spectrum is fitted with two Lorentzian functions and a linear background; the dashed and solid red curves denote the individual components and total fit, respectively. The two components overlap near the nodal direction but separate progressively towards the off-nodal directions. The extracted peak separation $|\Delta(\theta)|=|E_2(\theta)-E_1(\theta)|$ is plotted in polar form in Fig.~\ref{fig:polar}(i). 
To compensate for the intensity reduction imposed by the magnetic form factor, we construct the polar map by combining symmetry-equivalent sections of the high-energy magnon branches from neighbouring Brillouin zones.
The LSWT calculation (solid blue curve) reproduces the measured angular modulation, with maxima along the off-nodal directions and strong suppression near the nodes. This characteristic nodal pattern demonstrates that the observed magnon splitting is not simply a local band separation but a momentum-space manifestation of the underlying $g$-wave altermagnetic symmetry.\\

\noindent\textbf{Discussion}\\
To our knowledge, the present work is the first quantitative mapping of the in-plane momentum structure of the altermagnetic order parameter in hematite.
Beyond INS, complementary altermagnetic signatures in hematite have been reported using X-ray magnetic circular dichroism (XMCD) imaging~\cite{ishii_2026_fe2O3} and magneto-optical measurements~\cite{pan_2025_fe2o3}. However, because of the insulating nature of hematite, direct observation of spin-driven electronic band splitting using ARPES is challenging and has not yet been reported.

Circular-dichroic resonant inelastic X-ray scattering (CD-RIXS) has been used to identify altermagnet candidates, including MnTe~\cite{takegami_2025_Circular}, CrSb~\cite{biniskos_2025_Systematic}, La$_2$O$_3$Mn$_2$Se$_2$~\cite{zhang_d-wave_2026}, and Fe$_2$Mo$_3$O$_8$~\cite{channagowdra_2025_Fe2Mo3O8}. However, interpreting the dichroism as evidence of chiral split magnons remains controversial. RIXS circular dichroism can also arise from extrinsic optical effects, including X-ray birefringence and effective time-reversal-symmetry breaking inherent to the RIXS process~\cite{furo_2025_CDRIXS,nag_2025_birefringence}.
The momentum-dependent magnon splitting is instead a direct manifestation of the altermagnetic order parameter symmetry. By resolving the in-plane angular dependence of the splitting $\Delta(\theta)$ in hematite, we establish the $g$-wave symmetry of the altermagnetic state.
A key conclusion from our analysis is that the observed momentum-dependent splitting cannot be understood simply as a conventional magnon splitting caused by relativistic anisotropy \cite{verbeek_2024_fe2o3, hoyer_2025_fe2o3}, or dipolar interactions \cite{sears_2026_FeF2, Faure_2025_mnf2}. Single-ion anisotropy and Dzyaloshinskii–Moriya interactions can modify the magnon spectrum but do not account for the observed $g$-wave-like angular dependence of the optical branch splitting.
INS therefore provides a direct bulk probe of order parameter symmetry across a broad range of altermagnets, particularly in insulating systems where electronic band spin splitting is difficult to access by ARPES.

INS has been applied to several altermagnet candidates. In MnTe, magnon splitting was resolved along selected momentum trajectories, and comparison between nodal and off-nodal directions provided evidence for $g$-wave altermagnetic magnons~\cite{Liu2024Chirala}. More recently, INS measurements on single-crystal CrSb revealed a sixfold modulation of the constant-energy magnon contours, with a finite momentum-space separation along off-nodal directions and no corresponding separation along the nodal direction~\cite{singh_2026_crsb}.
Earlier INS measurements on $\alpha$-Fe$_2$O$_3$ observed the splitting along an off-nodal direction but did not determine its in-plane pattern~\cite{sun_observation_2025}. 
Nevertheless, the angular dependence of the magnon splitting has not yet been mapped quantitatively in any candidate material.
In this study, the high-quality data spanning broad reciprocal space enable such a map, establishing the $g$-wave altermagnetic order parameter in hematite.

The high rotational symmetry of a pristine $g$-wave state can suppress momentum-integrated linear spin response, although higher-order nonlinear spin current remains allowed~\cite{ezawa_2025_AM_nonlinear}. However, this constraint may be lifted through symmetry engineering. Recently, \textit{in situ} strain applied to free-standing hematite membranes was shown to control both the axial and basal-plane anisotropies, with anisotropic strain reducing the basal-plane symmetry and redistributing the antiferromagnetic domain populations~\cite{harrison_2025_fe2o3}. In CrSb, strain drives a $g$- to $d$-wave conversion and activates a sizable spin-splitter response~\cite{song_2026_crsb}, while crystal-symmetry reconstruction in thin films has enabled room-temperature anomalous Hall readout and electrical switching of the altermagnetic order~\cite{karetta_2025_theory,song_2026_crsb,zhou_2025_CrSb_film}. These results suggest that strain-induced changes in the altermagnetic magnon symmetry could be tested in hematite films or even bulk crystals.

In the context of spintronics, the insulating character of $\alpha$-Fe$_2$O$_3$ suppresses Ohmic dissipation, while its sharp and well-defined magnon modes demonstrate weak damping. These properties, together with hematite's high magnetic ordering temperature, large magnon energy scale, chemical stability, and availability in large, high-quality single-crystal form, make hematite a promising platform for realizing low-loss altermagnetic magnonics. Future room-temperature INS measurements could determine whether the angular dependence $\Delta(\theta)$ and its nodal structure persist across the Morin transition. Polarized INS could distinguish the $S^{+-}$ and $S^{-+}$ channels and probe the opposite handedness and domain dependence of the split magnons~\cite{liu_2026_mnte}. Together, these measurements would test the robustness and controllability of $g$-wave chiral magnons under technologically relevant conditions.\\

\noindent\textbf{Methods} \\
\noindent\textbf{INS experiments}\\
\noindent Inelastic neutron scattering experiments were performed on the HD time-of-flight spectrometer at the China Spallation Neutron Source (CSNS)~\cite{luo_2023_csns_hd}, using a 16.1 g hematite single crystal. The high crystallinity of the sample was confirmed by X-ray Laue diffraction (see Supplementary Information). The sample was aligned in the scattering plane spanned by the $(H,H,0)$ and $(0,0,L)$ directions and rotated within this plane in $1^\circ$ steps. The incident neutron energy was set to $E_i=129.5$~meV. The wave vector $\mathbf{Q}=(H,K,L)$ is indexed in reciprocal lattice units (r.l.u.) using the hexagonal unit cell. All measurements were conducted at 3.5~K.\\

\noindent\textbf{LSWT calculations}\\
\noindent Linear spin-wave theory calculations were performed using the SpinW package~\cite{toth_2015_spinw}. 
The collinear AFM spin structure with $S=5/2$ was adopted for all calculations (see Fig.~\ref{fig:overall}(a)).
To account for crystallographic twinning associated with the hexagonal lattice symmetry, contributions from the corresponding twin domains were considered in the calculations. Except for the symmetry-inequivalent exchange coupling pair $J_{13}^{(1)}$ and $J_{13}^{(2)}$, exchange couplings beyond $J_5$ were not included because their fitted values were small and did not improve agreement with the data.
\\

\noindent\textbf{Acknowledgments}\\
The work at CUHK is supported by the Research Grants Council of Hong Kong (CUHK 24306223, CUHK 14305926), the Guangdong Provincial Quantum Science Strategic Initiative (GDZX2401012), CUHK Direct Grant (4053671, 4053791), and the 1+1+1 CUHK-CUHK(SZ)-GDSTC Joint Collaboration Fund (2025A0505000079).
Y.F. is supported by Guangdong Basic and Applied Basic Research Foundation (Dongguan Joint Fund, Youth Fund Project, Grant No. 2025A1515110349).
X.T. acknowledges National Science Fund for Distinguished Young Scholars (Grant No. 12425512). \\

\noindent\textbf{Author contributions}\\
Q.W. and J.C. conceived the project.
T.W., A.L., Y.W., S.A. and J.C. prepared and characterized the sample.
T.W., Y.H.L., Y.W., Q.Z., Y.Y., W.L., X.T., Y.F. and Q.W. carried out the INS experiments. T.W., Y.H.L. and Q.W. analysed the data with assistance from Y.L. and Y.S.
T.W. and Y.H.L. performed LSWT calculations.
T.W. and Q.W. wrote the manuscript with input from other authors.\\

\noindent\textbf{Data availability}\\
Data supporting the findings of this study are available from the corresponding authors upon request. \\
	
\noindent\textbf{Competing interests} \\
The authors declare no competing interests.\\

\end{document}